\documentstyle[aps,pre,twocolumn,epsfig]{revtex}
\newcommand{\bel}[1]{\begin{equation}\label{#1}}

\def\bbbc{{\mathchoice {\setbox0=\hbox{$\displaystyle\rm C$}\hbox{\hbox
to0pt{\kern0.4\wd0\vrule height0.9\ht0\hss}\box0}}
{\setbox0=\hbox{$\textstyle\rm C$}\hbox{\hbox
to0pt{\kern0.4\wd0\vrule height0.9\ht0\hss}\box0}}
{\setbox0=\hbox{$\scriptstyle\rm C$}\hbox{\hbox
to0pt{\kern0.4\wd0\vrule height0.9\ht0\hss}\box0}}
{\setbox0=\hbox{$\scriptscriptstyle\rm C$}\hbox{\hbox
to0pt{\kern0.4\wd0\vrule height0.9\ht0\hss}\box0}}}}
\def\be{\begin{equation}}
\def\ee{\end{equation}}
\def\bege{\begin{equation}}
\def\ende{\end{equation}}
\def\bea{\begin{eqnarray}}
\def\eea{\end{eqnarray}}
\def\ba{\begin{array}}
\def\ea{\end{array}}

\def\si{\sigma}
\def\al{\alpha}

\def\eps{\epsilon}
\def\ra{\rightarrow}
\draft
\begin{document}
\tighten
\onecolumn

\title{ Symmetry breaking and phase coexistence in a driven diffusive
two-channel system }
\date{\today}
\author{Vladislav Popkov$^{1,2}$ and Ingo Peschel $^{1}$ }
\address{$^1$ Fachbereich Physik, Freie Universit\"at Berlin, Arnimallee 14,
D-14195 Berlin, Germany\\
$^2$ Institute for Low Temperature Physics, 310164 Kharkov, Ukraine\\
}
\maketitle
\begin{abstract}
We consider classical hard-core particles moving on two parallel
chains in the same direction. An interaction between the channels is
included via the hopping rates.
For a ring, the stationary state has a product form. For the case of coupling
to two reservoirs, it is investigated analytically and numerically. In
addition to the known one-channel phases, two new regions are found, in
particular
one, where the total density is fixed, but the filling of the individual chains
changes back and forth, with a preference for strongly different densities.
The corresponding probability distribution is determined and shown to have
a universal form. The phase diagram and general aspects of the problem
are discussed.

\end{abstract}

\section{Introduction}
\label{Intro}

Driven many-particle systems have been the topic of numerous studies in recent
years \cite{Schmitt}.  The simplest example is a one-dimensional lattice gas
where hard-core particles move stochastically in one direction. This model,
also
known as asymmetric exclusion process, can be treated exactly and therefore has
become a reference system in this field
\cite{Schuetz93,Derrida9398,Rezakhanlou91}.
The characteristic feature of such driven systems is a nonzero current in the
stationary state.  For the case of open boundaries, this current transports the
'information' from the boundaries into the bulk and nonequilibrium phase
transitions can arise \cite{Krug91,Mukamel95}.  These do not have analogues in
equilibrium systems where the boundaries normally do not play a significant
r\^ole.  It was shown in \cite{Popkov99} that the phase transitions for a
generic
driven system with one type of particles are governed entirely by an extremal
principle for the macroscopic current $j(\rho)$, where $\rho$ denotes the
average density of the particles.  This principle states that the stationary
bulk current assumes its absolute minimum within the interval set by the
boundary densities, if $\rho_-$(left) $< \rho_+$(right) and the flux is towards
the
right. If $\rho_-> \rho_+$ it takes its absolute maximum.
In either case, the current and the density fluctuate only slightly
around their stationary values and the fluctuations vanish as the system size
goes to infinity.

In the present communication we study a system with two parallel
channels and show that these two properties do not hold in general.
Our example is a simple extension of the one-channel model and contains
one additional parameter which measures the coupling between the channels.
There is no exchange of particles, but the hopping rates in one chain depend
on the local configuration in the other one.
They are chosen in such a way that, for a ring, the stationary state has
a simple product form and thus the current can be obtained explicitly.
With reservoirs at the ends, the system was studied by a combination of
numerical and analytical methods. In a $\rho_-$ - $\rho_+$ phase diagram,
it shows a number of regions with equal densities in both chains and various
values for the current. In addition, however, there are two other regions
with unexpected new features. They develop out of the first-order transition
line of the single-chain problem as the interaction between the channels is
turned on. In both of them, the overall densities $\rho_1$ and $\rho_2$ in
the two chains fluctuate
strongly and only a probability distribution $w(\rho_1,\rho_2)$ can be given.
Together with that, symmetry-breaking phenomena appear.

In one region, one finds spatial coexistence between sections of equal
and of unequal (but fixed) densities in the two chains, with the size of these
sections varying in time. This is similar to the situation on the transition
line for one chain, where sections of high and low density, separated by
a domain wall, coexist. It is related to the fact that the current, which must
be constant throughout the system (and thus plays the r\^ole of a chemical
potential),
can be the same for different densities. This will be called the {\it
mixed-phase
region} in the following.

In the other region, which appears if the coupling exceeds a critical value and
then
grows at the expense of the first one, only the unsymmetric configurations
exist.
The total system is then practically half filled, $\rho_1 + \rho_2 =1$, but the
individual densities change in time. The same holds for the current.
The most probable configurations are those where
one channel is relatively empty and the other relatively full. Between them
the system diffuses back and forth and hence we will use the term {\it seesaw
region}
in the following. The time in which the channels interchange
r\^oles increases only as a power of the system size, in contrast to the result
for
a model with two kinds of particles on one chain
\cite{Mukamel95,Rittenberg-chain}
or simplified versions of it \cite{Mukamel_walker,Arndt-walker}.
The situation can be compared to that at an equilibrium
first-order transition with a vanishing or size-independent free-energy barrier
between
the phases. In this sense, the symmetry breaking could be called `weak'. The
origin
of this behaviour is related to the existence of fast and slow processes in the
system,
as will be discussed in detail.

The paper is organized as follows. In section II we define the model and
describe the
solution on a ring. In section III we treat it with boundary reservoirs and
present
numerical results from Monte-Carlo and mean-field calculations. The case of
strong
interactions, where one can simplify the problem and obtain analytical results
for
the seesaw region, is the topic of section IV. After that we turn in section V
to general
interactions and establish the complete $\rho_- - \rho_+$  phase diagram.
Section VI
contains the conclusion and some discussion of open problems. Some details
concerning
the boundary rates and the mean-field equations can be found in Appendices A
and B.

\section{Model}
\label{Model}

Our model consists of two parallel chains, on each of which particles
can hop towards the right if the next site is empty. The
hopping rate in one chain depends on the configuration of the
neighbouring sites in the other and one has the four processes
shown in Fig.~\ref{fig_process}. The stationary state has a simple form (given below) 
if the
rates satisfy the condition $\alpha + \beta = 2 \gamma$. In the
following we choose $\alpha = \beta = \gamma = 1$, so that the
remaining rate
\be
\epsilon \equiv exp(-\nu)
\ee
is the only parameter.
It refers to those processes where the site besides the jumping
particle is empty, but the next one in the forward direction is
occupied. We will always assume $\epsilon < 1$.  This can be
viewed as the result of short-range interactions which increase
the barrier similar as in \cite{Singer80}. In a traffic-model
context, $\epsilon$ would describe a hesitation to move besides
another car.

\setlength{\unitlength}{1.8cm}
\begin{figure}[!h]  
\begin{center}
\epsfig{width=7\unitlength,
       angle =0,
      file=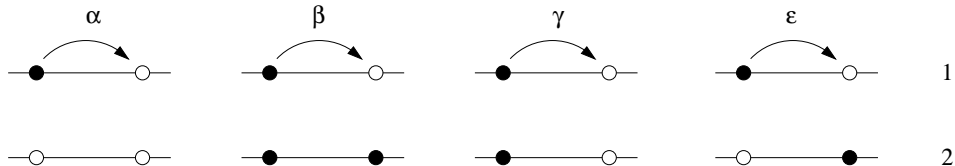}\vspace{3mm}
\caption{ The four elementary hopping processes, shown here for the first
chain,
          and their rates. In the study, the first three rates are set equal to
1.}
\label{fig_process}
\end{center}
\end{figure}

\begin{figure}
\begin{center}
\epsfig{width=7\unitlength,
       angle =0,
     file=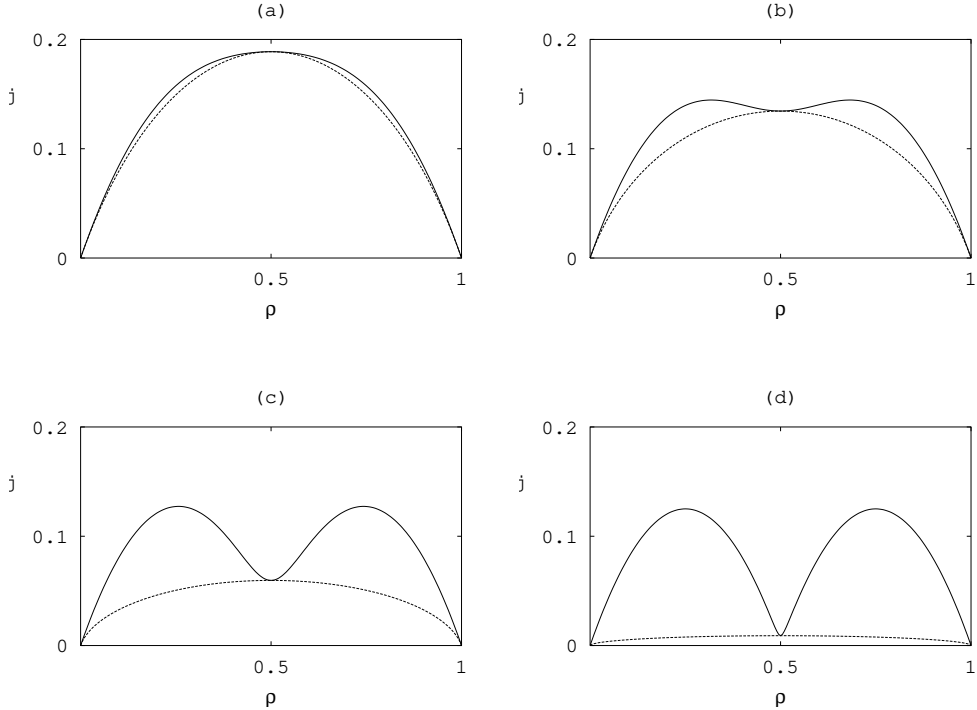}\vspace{3mm}
\caption{ Current $j$ vs. density $\rho$ for a ring according to (3,4) for four
         values of the interaction, $\nu=1.2, 2, 4, 8$ (a-d). Upper and lower
curves
         correspond to $\rho = \rho_1 = \rho_2$ and $\rho = \rho_1 = 1
-\rho_2$,
         respectively.}
\label{fig_flux}
\end{center}
\end{figure}

The state of the system can be described either by occupation numbers
$n_k,m_k$  or by spin variables  $\sigma_k, \tau_k$ for the two chains.
The first take the values 1(0), the latter the values +1(-1) if site
$k$ is occupied (empty).
For a ring with $N$ sites, the stationary probability $P(\sigma,\tau) =
P(\sigma_1,\sigma_2, \ldots ,\sigma_N,\tau_1,\tau_2, \ldots ,\tau_N)$
then has the Boltzmann form

\be
P(\sigma,\tau) = C
\prod_{k=1}^N exp(- {1 \over 4} \nu \si_k \tau_k)
 \label{stationary}
\ee
which means that different sites $k$ are uncorrelated,
This can be proved by a straightforward consideration of the
gain and loss processes. For the ring geometry, the stationary
densities are constant and the quantities of main interest are
the currents $j_1,j_2$ in the two chains. These can be calculated
either by working in a grand canonical ensemble or via the
mean-field equations of Appendix B which are exact here due to
the form of (\ref{stationary}). For example, the mean-field expression for
$j_1$ is, from (\ref{n_k}), with $p_k = n_k m_k$

\be
j_1 =
n_k (1-n_{k+1}) + (\epsilon-1)(n_k-p_k)(m_{k+1}-p_{k+1})
\label{j1}
\ee
where all quantities are expectation values. In the
stationary state they are independent of $k$, $n_k = \rho_1, m_k = \rho_2$
and one only has to determine $p$. The expression for the current for arbitrary
densities 
$\rho_1, \rho_2$ is complicated and is omitted here. In particular cases
 $\rho_1=\rho_2$ or
$\rho_1=1-\rho_2$, the currents in the two chains are the same and given
by

\begin{eqnarray}
&j =  \rho \left( 1-   \rho \right)
\left[
1 \pm
{\left( 1- \sqrt{1+F(\pm\nu,\rho)}\right)^2 /F(\pm\nu,\rho) }
\right];
\label{j_theor}   \\
&F(\nu,\rho) = 4 \left( e^{-\nu} -1 \right)
\rho  \left( 1-   \rho \right),
\end{eqnarray}
where the upper (lower) sign holds for $\rho_1=\rho_2=\rho$ and
$\rho_1=1-\rho_2=\rho$, respectively. The resulting curves are shown in
Fig. \ref{fig_flux} for various values of the interaction parameter $\nu$. One
can see
that
for larger $\nu$ a minimum at $\rho=1/2$ exists in the case of equal
densities.
The value where this first happens is $\nu_{crit} = ln4 \approx 1.39$.
The reason is that for half filling and large $\nu$ adjacent
sites are mainly occupied by particle-hole pairs and then
only hopping with the small rate $\epsilon$ is possible.
Such a double-peak structure of the current
leads to a much richer phase diagram in the one-chain problem with
reservoirs
\cite{Popkov99} and will be important here, too.


\setlength{\unitlength}{1.9cm}

With reservoirs, particles enter the system at site $n=1$ and leave from
site $N$. If one adds reservoir sites $n=0$ and $n=N+1$ to the chains, the
bulk processes of Fig.~\ref{fig_process} appear also at the boundaries. The rates can then
be chosen in such a way that the dynamics is the same as in the interior,
but at prescribed densities $\rho_-$ and $\rho_+$ \cite{Antal00,Popkov99}.
These are thus the only boundary parameters which enter. The procedure is
described in more detail in Appendix A. In the following we will normally
choose equal boundary densities for both chains which makes the problem
completely symmetric between them. Since the transport can
be viewed as the motion of vacancies towards the left, it is also
symmetric under the exchange $\rho_- \leftrightarrow 1-\rho_+,
\rho_{\alpha} \leftrightarrow 1-\rho_{\alpha}$.

\section{numerics}
\label{numerics}

We have performed Monte-Carlo (MC) simulations of our system
for different values of $\rho_-,\rho_+$ and $\nu$, looking first at
global quantities, namely the overall densities $\rho_1(t)$ and
$\rho_2(t)$ in the two chains. The motivation for this
was that, for a single chain, the average density of particles
is an order parameter characterizing the different phases. This means that
$\rho(t)$ fluctuates only slightly around its mean value and the fluctuations
vanish as the system size grows.

\begin{figure}
 \begin{center}
\epsfig{width=7\unitlength,
       angle =0,
      file=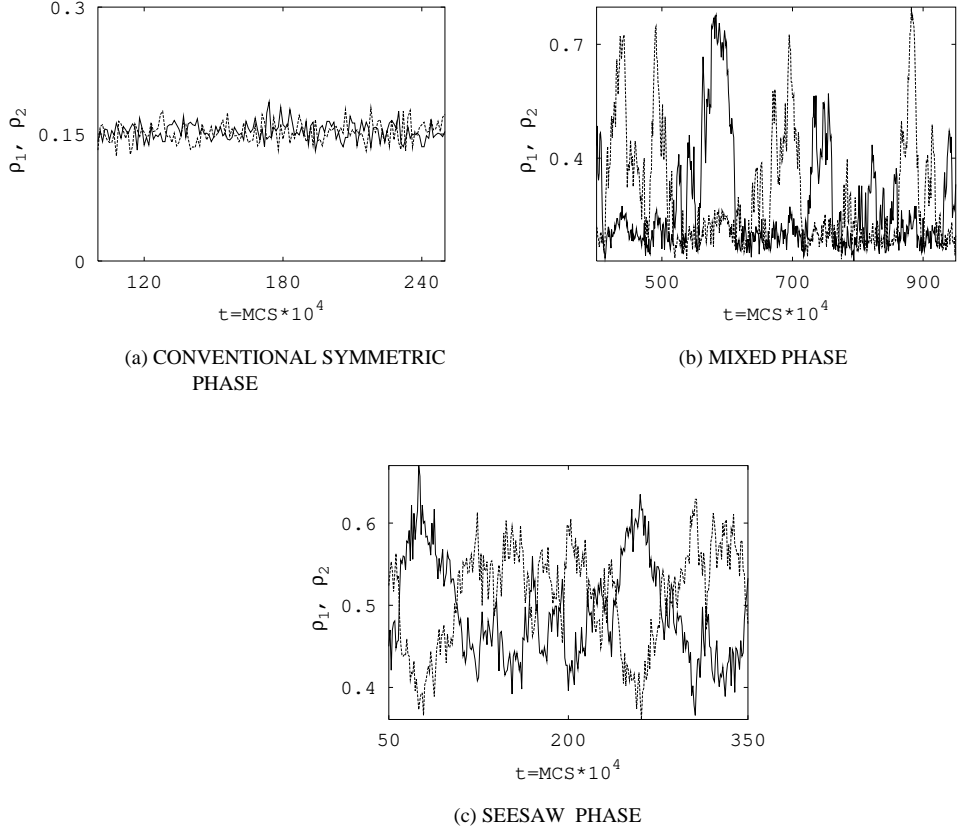}\vspace{3mm}
\caption{ Time evolution of the average densities in the two chains for
 three typical cases, as obtained from simulations for a system with $\nu = 2$
 and $N = 1000$ sites.
 Boundary densities $\rho_-,\rho_+$ : (a) 0.15, 0.5 ; (b) 0.15, 0.8 ;
 (c) 0.42, 0.6. }
\label{fig_evolution}
\end{center}
\end{figure}

Typical results of such MC calculations are presented in
Fig.\ref{fig_evolution}
for $\nu = 2$ and three different values of the boundary densities.
While in Fig.\ref{fig_evolution}(a) the behaviour
is as for a single chain, the other two figures show a `random walk'
of $\rho_1$ and $\rho_2$ within a large range of densities. These
large fluctuations are intrinsic, they do not change
qualitatively with the system size.
In Fig.\ref{fig_evolution}(c) the situation is still relatively
simple since the sum of the densities stays approximately
constant. In Fig.\ref{fig_evolution}(b),
however, the behaviour is
rather irregular and the global densities do not give a sufficient
description. Rather one has to look at the state of the system in
more detail. It then turns out that the seemingly chaotic
pattern is connected with the spatial coexistence of several phases.


\setlength{\unitlength}{1.9cm}

\begin{figure}
 \begin{center}
\epsfig{width=7\unitlength,
       angle =0,
      file=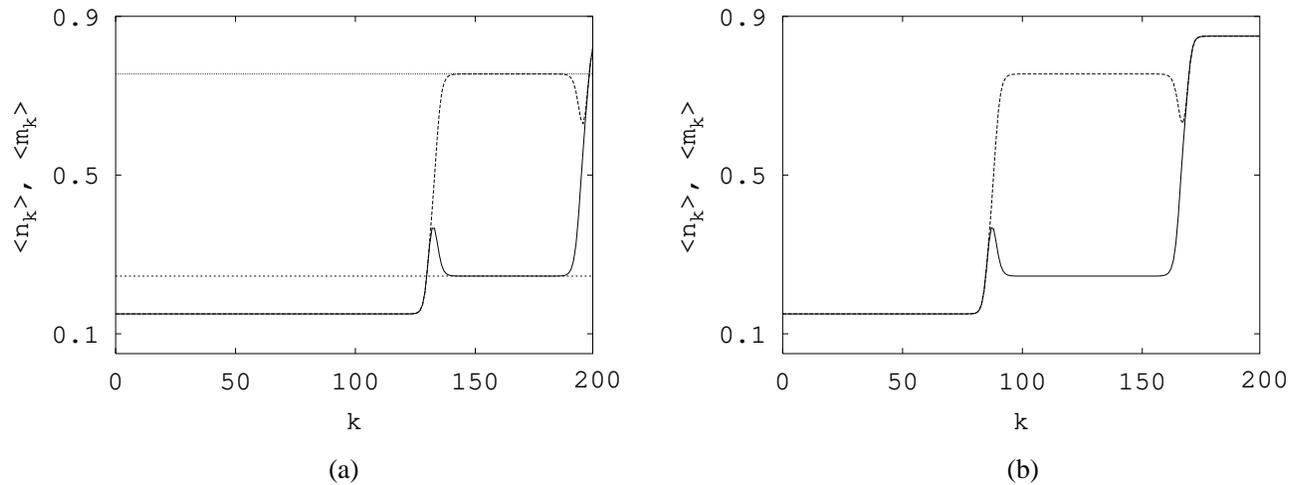}\vspace{3mm}
\caption{ Density profiles in the mean-field approximation for a system
 with $\nu = 2$ and $N=200$  sites. Boundary densities $\rho_-, \rho_+$ :
(a) 0.15, 0.82; (b) 0.15, 0.85. The horizontal
lines indicate the densities  $\hat\rho, 1-\hat\rho$ predicted by equating the
currents
 $j(\hat \rho,1-\hat \rho)=j(\rho_-,\rho_-)$, see section \ref{numerics}.}
\label{fig_MF}
\end{center}
\end{figure}

This can already be seen in a mean-field (MF) analysis of the system.
The corresponding equations are given in Appendix B. We integrated them over
time,
starting from random initial conditions and stopping
when the current had converged to $10^{-6}$ in
relative units. In the region of small $\rho_-$ and
large $\rho_+$, the mean-field density profiles then
had the typical shapes seen in Fig.\ref{fig_MF}.

\setlength{\unitlength}{2.5cm}

Let us first discuss Fig.\ref{fig_MF}(a) which corresponds to
$\rho_- < 1-\rho_+$ . Here one has two regions where the local densities
are constant. On the left they are the same in both chains,
$\langle n_k\rangle,\langle m_k\rangle =\rho_- $, while
on the right they are different. This leads to the bubble-like
structure in the figure. That such a coexistence is possible,
follows already from Fig.\ref{fig_flux}, since a given value of the current
(for the ring)
can be realized in different ways. Correspondingly, the densities
in the bubble region are given by
$\langle n_k\rangle = \hat \rho,\langle m_k\rangle = 1-\hat \rho$
where $j(\hat \rho,1-\hat \rho)=j(\rho_-,\rho_-)$.
The left end of the bubble can be anywhere, its location depends on the initial
conditions. Similarly, the two channels can exchange roles.
However, the bubble is always `glued' to the right boundary.
For $\rho_- > 1-\rho_+$
one has a similar picture but
the region with equal densities now appears on the right and the
values are $\langle n_k\rangle,\langle m_k\rangle =\rho_+ $.
This is a consequence of the particle-hole symmetry of the model.
The bubble in this case is attached to the left boundary, while the
location of its right end depends on the initial conditions.
Finally, on the line $\rho_-=1-\rho_+$, see Fig.\ref{fig_MF}(b),
the bubble coexists with {\it two} regions, one to the left and one
to the right, where the density in both chains is the same. On the left
its value is $\rho_-$, on the right it is $\rho_+$. These regions can coexist,
because $j(\rho,\rho)= j(1-\rho,1-\rho)$. In this case, the locations of
both ends of the bubble depend on the initial conditions.

In order to verify the correctness of this MF picture,
we performed specific Monte-Carlo simulations. In the spirit
of Derrida et al. \cite{Derrida}, we introduced two phantom
particles $A$ and $B$ designed to track down
the left and right end of a bubble, respectively.
Denoting the position of particle $A$ by $a$, the rules are:

\begin{eqnarray}
  \label{rulesA1}
  a \rightarrow a +1, \mbox{ if }&  n_{a +1}=0, m_{a +1}=0 \\
  a \rightarrow a -1, \mbox{ if }&  (1-n_{a -1})(1-m_{a -1}) =0
\end{eqnarray}

This means that $A$ moves preferentially to the right
in a low-density region and preferentially to the left
in a high density region, including the bubble region.
In contrast to the second-class particles in Ref.\cite{Derrida},
$A$ does not use the sites of the chains and should rather
be viewed as moving besides them. Its dynamics therefore
does not interfere with the that of the chain particles.
Analogously, the dynamics of particle $B$ with coordinate $b$ is

\begin{eqnarray}
  \label{rulesB1}
  b \rightarrow b -1,  \mbox{ if } &  n_{b -1}=1, m_{b -1}=1 \\
  b \rightarrow b +1,  \mbox{ if } &   n_{b -1} m_{b -1} =0
\end{eqnarray}

and it therefore tracks the right end of a bubble.

\begin{figure}
 \begin{center}
\epsfig{width=7\unitlength,
       angle =0,
      file=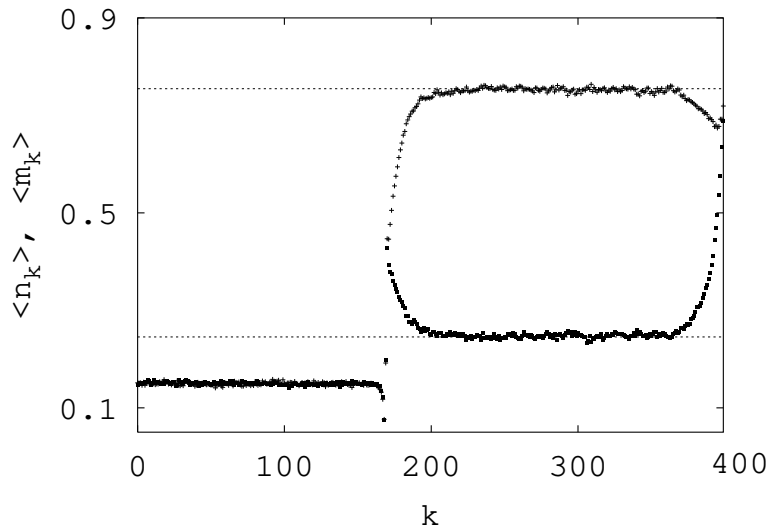}\vspace{3mm}
\caption{ Average density profiles as seen from a second-class particle
positioned at site 170, provided  the density in the first chain is higher,
for a system with $\nu$=2, $\rho_- = 0.15, \rho_+ = 0.77$ and $N =400$ sites.
The average is taken over
$6*10^7$ Monte Carlo steps after $3*10^5$ steps of equilibration. The
horizontal
lines indicate the densities  $\hat\rho, 1-\hat\rho$ predicted by equating the
currents
 $j(\hat \rho,1-\hat \rho)=j(\rho_-,\rho_-)$, see section \ref{numerics}.}
\label{fig_MC1}
\end{center}
\end{figure}

One then runs the MC simulations, adds particles $A$ and $B$ at
some point and monitors their positions.
For $\rho_- < 1-\rho_+$ it turns out that $A$ performs a random walk,
while $B$ basically sticks to the right boundary, $b \approx N$.
It is then interesting to look at those configurations,
where $A$ is at a certain specified site. Thereby one singles out the
states with a particular size of the bubble region (if there is one).
The average density profile in this case is shown in Fig.\ref{fig_MC1}
for $\rho_- = 0.15, \rho_+ = 0.77$.

\setlength{\unitlength}{1.5cm}

Quite remarkably, it has the same shape as found in the
mean-field calculations \cite{local_decrease}. This confirms that, indeed,
there is a very unusual dynamical coexistence of various states.
The MC calculations also confirm that the size of the bubble region
can vary and that the two chains can interchange roles. While this
is related to the initial conditions in the MF treatment, it happens
dynamically here.

For the case of Fig. \ref{fig_evolution}(c) the situation is different.
As mentioned, the system is always half-filled, i.e.
$\rho_1(t)+\rho_2(t)=1$ during the evolution. For given overall densities
$\rho_1, \rho_2$, one finds an average profile with
one bubble which fills essentially the whole
system. Thus there is no spatial coexistence of different regions along
the chains. However, the densities individually fluctuate strongly in
an interval
\be
\rho_{min} \leq \rho_{1,2} \leq \rho_{max}
\label{seesaw_boundaries}
\ee
where roughly $\rho_{min} = \max(\rho_-,1-\rho_+),\  \rho_{max} =
1-\rho_{min}$.
A small asymmetry $h$ in the boundary densities (a field in magnetic language)
suppresses these fluctuations. If one chooses
\be
\rho_{-,1} = \rho_- - h,\;\;\rho_{-,2} = \rho_-,\;\; \rho_{+,1} = \rho_+,\;\;
 \rho_{+,2} = \rho_+ + h;
\label{field}
\ee
which makes the chains inequivalent without destroying the particle-hole
symmetry,
the system locks in at $\rho_1 =\rho_{min},\  \rho_2 =\rho_{max}$ or vice
versa,
depending on the sign of $h$ \cite{Rittenberg_thanks}.
Moreover, if one computes formally a 'free energy density'
$f(M) = - log[w(M)]/N$ at small field, where
$w(M)$ is the stationary probability to have $M$ particles in one channel,
it has exactly the same form as found for a single chain in the vicinity of
the line of first-order transitions, see Fig. 18 in \cite{Rittenberg-chain}.
Therefore, the whole region where this occurs (and which we will determine in
more
detail below) can be viewed as one of first-order transitions.

\section{Strong-interaction limit}

The behaviour described at the end of the previous section is found in the
whole domain of reservoir densities
\be
\rho_- <1/2;\ \ \rho_+>1/2
\label{ddp_domain}
\ee
if the interaction between the channels is strong, i.e. if $\nu \gg 1$
($\epsilon \ll 1$).
In this limit, the problem can be discussed analytically, as we now show.

\begin{figure}[!h]  
 \begin{center}
\epsfig{width=7\unitlength,
       angle =0,
      file=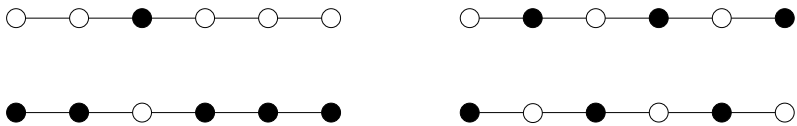}\vspace{3mm}
\caption{Two examples of metastable states, see text.}
\label{fig_configurations}
 \end{center}
\end{figure}

The basic observation is that in this case {\it three} of the boundary
processes
are slow (with rate of order $\epsilon$) and only {\it one} (where on the left
both sites are empty or on the right both sites are full) is fast (rate
of order 1). This comes from the construction of the boundary rates
mentioned above and is discussed in more detail in Appendix A.
As a consequence, there exist particular configurations,
which can be left only with rate $\epsilon$, and which we will call
metastable. These are such that the system is half filled, with $M$
particles in the first and $N-M$ in the second channel, arranged in
such a way that adjacent sites are always in a particle-hole state.
Two examples are shown in Fig.~\ref{fig_configurations}.

\setlength{\unitlength}{1.5cm}

Transitions between 
such states
have the character of avalanches. After a slow process, either at the
boundary or in the interior, a sequence of $N$ fast processes follows
after which a new metastable configuration is reached. If the start
is in the bulk, four different final configurations with $M' = M,M \pm 1$
can be reached according to
\be
M \ra
\left\{
\begin{array}{ccc}
  M  & \mbox{with probability } & 1/2  \\
  {M+1}  & \mbox{with probability } & 1/4  \\
  {M-1}  & \mbox{with probability } & 1/4
  \end{array}
\right.
\label{bulk}
\ee


A process in the bulk starts at a wall, where the orientation of the
particle-hole pairs changes. For example, the first configuration in
Fig.~\ref{fig_configurations} has two such walls. For a given $M$, the average number of walls
can be shown to be $\sim M (1-M/N)$. Therefore the average rate for
bulk processes from level $M$ is given by
\be
\alpha_M \sim M (1-M/N) \epsilon /4
\label{rate}
\ee
Similar considerations apply to the processes starting at the boundaries. Here
two final configurations $M' = M$ and $M' = M+1$ or $M' =M-1$ can be reached
with equal probability. However, the rate does not involve $M$ and thus these
processes are unimportant compared to the bulk processes except for very
small $M$ or $N-M$. In the following we will neglect them.

\begin{figure}
 \begin{center}
\epsfig{width=7\unitlength,
       angle =0,
      file=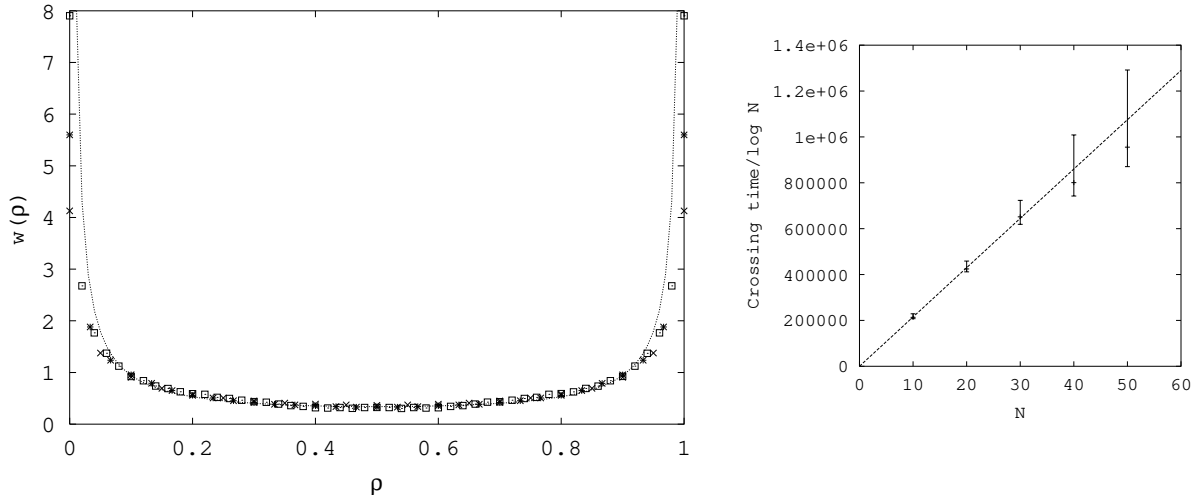}\vspace{3mm}
\caption{ (a) Stationary probability distribution $w(\rho)$ for small systems
in the seesaw region.
  Analytical result (11) (continuous curve) in comparison
  with data from Monte-Carlo simulations of systems with $\nu = 8, \rho_- =
0.1,
  \rho_+ = 0.9$ and $N=20,30,50$  sites (crosses, stars and boxes).
  The systems have evolved for at least
   $10^7$ Monte-Carlo steps, the averages are taken over 40 histories.
  (b) The corresponding average passage times.}

\label{fig_nu8}
 \end{center}
\end{figure}

The dynamics of the metastable states is thus equivalent to a random walk
in the space of occupation numbers $0 \le M \le N$ with a position-dependent
rate given by (\ref{rate}). This rate, which in terms of the density
$\rho_1 = \rho = M/N$ is proportional to $\rho (1-\rho)$,
is small at the ends of the interval and large in the middle.
Therefore the system will spend most of its
time near $M = 0$ and $M = N$. This can be made more precise by writing
down the master equation for the probability $w_M$ to find the value $M$
\be
\frac {\partial w_M } {\partial t} =
\alpha_{M+1} w_{M+1} + \alpha_{M-1} w_{M-1} - 2 \alpha_M w_M
\label{master}
\ee
In a continuum limit, this becomes
\be
\frac {\partial w(\rho) } {\partial t} =
\frac {{\partial}^2} {\partial {\rho}^2}
\left( D(\rho) w(\rho) \right)
\ee
where $D(\rho) = \alpha_M/N^2$ is the diffusion constant. This equation can
be related to the associated Legendre differential equation. However, for
the stationary state it is sufficient to note that the probability current
is given by
\be
I(\rho) = -\frac {\partial} {\partial {\rho}}
\left( D(\rho) w(\rho) \right)
\ee
Since this has to be zero, the stationary distribution follows as
\be
w(\rho) = \frac{A} {\rho (1-\rho)}; \ \ 0 < \rho < 1
\label{w}
\ee
This simple universal function, which is independent of the parameters
of the system, is also seen in simulations as illustrated in
Fig.~\ref{fig_nu8}(a).
Actually, the picture shows some deviations at the boundaries of
the interval, but there
the neglected processes should be included. This would lead to finite
rates also at $\rho = 0,1$ and prevent $w(\rho)$ from diverging there.
In order to normalize it as it stands, one has to leave out a small
boundary region of width $\delta/N$.

\setlength{\unitlength}{2.5cm}

Although the system is mainly in configurations with the density in one
chain small and in the other large, it is not locked into them. The average
time
in which the two chains interchange r\^oles can be calculated from
the first passage time formula given in \cite{van Kampen}. In the
continuum case it reads
\be
T = \int_{\delta/N}^{1-\delta/N}
 d \rho w(\rho) \int_{\delta/N}^{1/2} {d \rho' \over {D(\rho') w(\rho')}}
= N ln (N/\delta) /\epsilon
\label{T_cross}
\ee
Thus $T$ is smaller than for a homogeneous diffusion process
where it varies as $N^2$. This is due to the large hopping rates proportional
to $N$ for intermediate values of $\rho$. The law is also found numerically,
as seen from Fig.~\ref{fig_nu8}(b). However, this result and
the considerations so far do not apply to arbitrarily large system sizes.

If $N$ becomes too large, the typical time $1/N\epsilon$ after which a new slow
process starts, becomes smaller than the time $\tau = N$ needed to complete
the first one. Therefore the previous considerations are limited to sizes
such that $N^2\epsilon \lesssim 1$. For larger systems, it is not possible to
separate the fast and the slow processes. Nevertheless, the system is still
basically half filled and each time a particle comes in and another particle
exits, the number $M$ changes according to (\ref{bulk}). Since these processes
also determine the flux through the system, their effective rate can be related
to $j(\rho,1-\rho)$. Thus instead of (\ref{rate}) one has
\be
\alpha_M = j(\rho,1-\rho)/2
\label{rate-j}
\ee
where for $j$ one can use the ring result (\ref{j_theor}). This should give the
behaviour in the thermodynamic limit. Following the same steps as before,
one then obtains the distribution
\be
w(\rho) = \frac{A} {j(\rho,1-\rho)}; \ \  \rho \mbox{ within the limits
(\ref{seesaw_boundaries})   }
\label{winf}
\ee
for the density $\rho$. This is again independent of the boundary values
$\rho_\pm$, and the interaction enters only via the formula for the current.
The function is qualitatively similar to (\ref{w}) but,
according to the form of $j$ (cf. Fig.\ref{fig_flux}), has a flatter shape.

\setlength{\unitlength}{1.5cm}
\begin{figure}
 \begin{center}
\epsfig{width=7\unitlength,
       angle =0,
      file=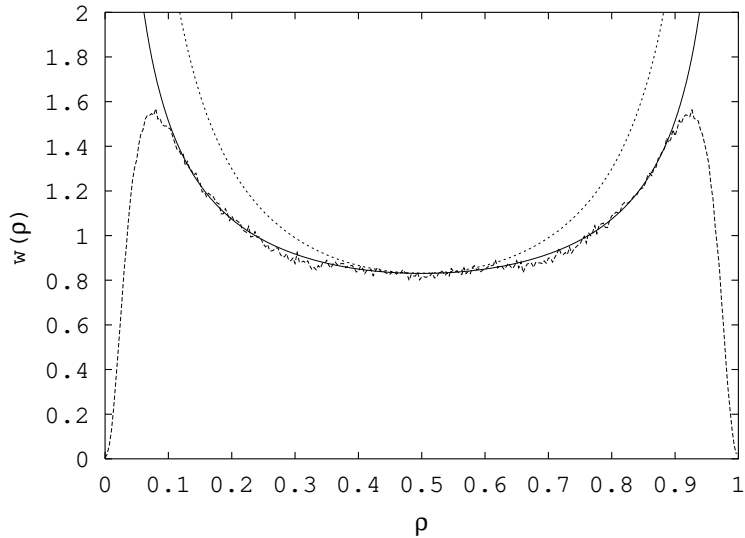}\vspace{3mm}
\caption{Stationary probability distribution $w(\rho)$ for a large system
         in the seesaw phase. Analytical result (14) together with
         with simulation data for a system of $N= 300$ sites from $1.5\cdot
10^7$
         Monte-Carlo steps, averaging over 40 histories. The parameters are
         $\nu = 4, \rho_- = 0.2$ and $\rho_+ =0.8$. The result (11)
         is also shown for comparison (dotted). }
\label{fig_thermodynamic_omega}
 \end{center}
\end{figure}

This behaviour
is clearly seen in simulations. Fig.\ref{fig_thermodynamic_omega}
shows numerical results for a system of
$300$ sites together with the analytical prediction. Again, there are
deviations
in the boundary regions near $\rho = 0$ and $\rho = 1$ due to
the Eq.\ref{seesaw_boundaries}
 but apart from that
the agreement is very good. The curve (\ref{w}),
plotted for comparison, clearly does not fit the data. Note that the rate
(\ref{rate-j}) does not contain the size any more. Therefore, the passage
time $T$ is now proportional to $N^2$. In contrast to the diffusion model
treated in \cite{Mukamel_walker,Arndt-walker} there is no exponential
increase with the size.

\section{Phase diagram}

The boundaries of the seesaw region discussed above, but also of the mixed
region, depend on the interaction parameter $\nu$ and can be found from the
following argument. Let us fix $\rho_+ > 1/2$
and gradually increase $\rho_-$. For very small $\rho_-$, it takes the time
$t \approx 1/\rho_-$ to fill an empty site from the reservoir, which is longer
than the time $1/\epsilon$ for slow processes to happen in the bulk. In this
case
the system will go into the symmetric low-density phase which also exists in
the lower left corner of the $\rho_- - \rho_+$ phase diagram. As $\rho_- <
\rho_+$, it will tend to minimize its flux according to the minimization
principle \cite{Popkov99} and the stationary current will be
$j(\rho_-,\rho_-)$.
But this principle, extended to the two-channel case, suggests that the low-
density phase will become unstable if there is a region $(\rho_1,\rho_2)$ such
that
the currents in both channels are smaller than $j(\rho_-,\rho_-)$ :
\be
j_\alpha(\rho_1,\rho_2) < j(\rho_-,\rho_-);\ \  \rho_- < \rho_1,\rho_2 < \rho_+
\label{bound}
\ee
A transition will take place as soon as the first such point appears.
For our system this happens at $\rho_1 = 1 - \rho_2 = \rho_+$ and therefore the
instability is expected when
\be
j(\rho_-,\rho_-) = j(\rho_+,1 - \rho_+)
\label{cond1}
\ee
i.e. when the upper and the lower curves in Fig.\ref{fig_flux}   lead to the same 
current.
Actually,
according to the results in section III, the low-density phase does not vanish
completely
but coexists with unsymmetric configurations beyond that point.

\begin{figure}
 \begin{center}
\epsfig{width=7\unitlength,
       angle =0,
      file=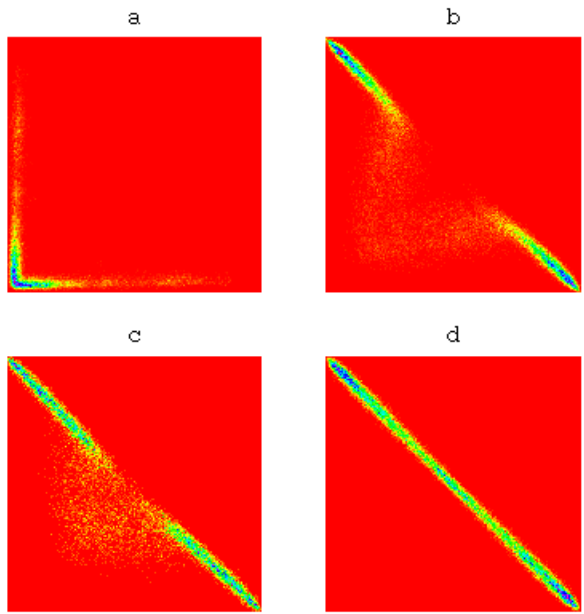}\vspace{3mm}
\caption{ Stationary distribution $w(\rho_1,\rho_2)$ as obtained
         from numerical simulations for a system of 300 sites with $\nu = 4$,
         $\rho_+ = 0.8$ and $\rho_- = 0.02, 0.055, 0.06, 0.09$ (a-d).
         The figures show the location of non-zero $w$,
         the dark regions within white ones correspond to the highest values of
$w$.}
\label{fig_contours}
 \end{center}
\end{figure}

The part of the region (\ref{bound}) along the diagonal $\rho_1 + \rho_2 = 1$
consists of two separated segments (compare
Figs.\ref{fig_contours}b,c).
As $\rho_-$ grows, these segments also grow until they finally merge
at a value $\rho^{*}$ such that
\be
j(\rho^{*},\rho^{*}) =j(1/2,1/2)
\label{cond2}
\ee

\begin{figure}
 \begin{center}
\epsfig{width=7\unitlength,
       angle =0,
      file=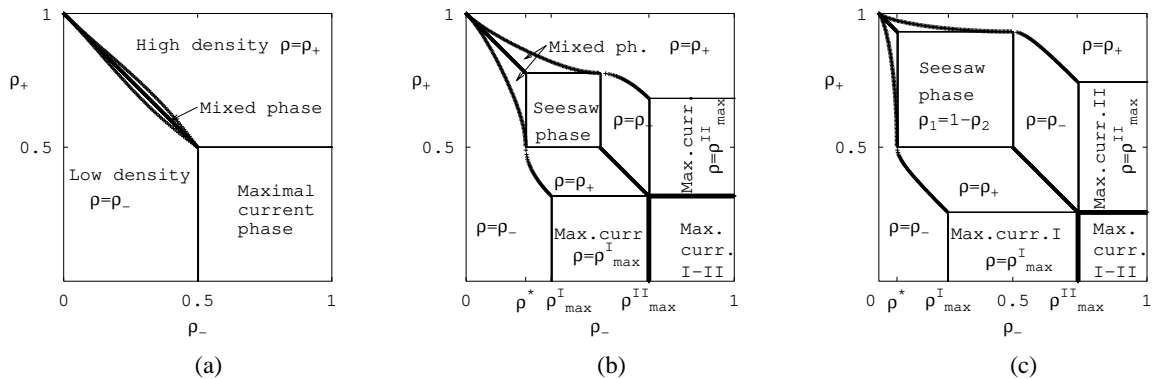}\vspace{3mm}
\caption{Phase diagram of the model for three different values of the coupling
:
$(a) \nu = 1$, $(b) \nu = 2$, $(c) \nu = 4$. Thick (thin) lines indicate
         first (second) order transitions. $\rho^I_{max}$ and $\rho^{II}_{max}$
denote
         the positions of the left and right maximum of the curve
$j(\rho,\rho)$
         in Fig. 2.}
\label{fig_phase_diagrams}
\end{center}
\end{figure}

From that point onwards, the low-density symmetric region
(the presence of which is signalled by the `cloudy' part in
 Figs.\ref{fig_contours}b,c)
disappears completely and the seesaw phase described in the previous
section takes over.
A further increase of $\rho_-$, however, decreases the range within which
$\rho_1$ and $\rho_2$, according to  (\ref{bound}), (\ref{seesaw_boundaries}),
may fluctuate. At $\rho_- = 1/2$ the system crosses the
boundary into the symmetric phase with $\rho_1 =\rho_2 = \rho_+$ via a second-order
transition
(see Figs.\ref{fig_phase_diagrams}b,c).

These simple arguments are well supported by Monte-Carlo calculations.
In Fig. \ref{fig_contours}
the stationary density distribution $w(\rho_1,\rho_2)$ is shown for fixed
$\rho_+$
and four different values of $\rho_-$. The change of its shape as one crosses
the points defined by (\ref{cond1},\ref{cond2}) is clearly seen.

\setlength{\unitlength}{1.2cm}

In Fig. \ref{fig_contours} (a) the system is in the conventional low-density
phase. The distribution
has a boomerang shape due to the finite size (incidentally similar to the one
found
in \cite{Rittenberg-chain}) with the weight however concentrated on a
symmetric line $\rho_1=\rho_2$ near the origin.

In Fig. \ref{fig_contours} (d), on the other hand, there is only weight along
the
diagonal $\rho_1+\rho_2=1$  which is
characteristic of the seesaw region. Figures \ref{fig_contours} (b) and
(c) show the transition region between (\ref{cond1}) and (\ref{cond2}) and
indicate
the mixed  phase. The `cloudy' part in Fig.\ref{fig_contours} (b) and
(c) is due to the coexistence of symmetric and asymmetric states mentioned
above
and the increasing weight along the diagonal comes from an increase of density
values satisfying Eq.(\ref{bound}).
The mixed region, as seen in Figs. \ref{fig_phase_diagrams}(a)-(c),
exists for all values $\epsilon <1$.
By contrast, the seesaw region only appears
when $\epsilon$ becomes smaller than $\epsilon_{crit}=1/4$, the value where the
symmetric current $j(\rho,\rho)$ starts to develop a double-hump structure.

The discussion so far has assumed that the value of $\rho_+$, for which one
varies
$\rho_-$, is not too large. If $\rho_+ > \rho^*$, where $\rho^*$ is defined by
(\ref{cond2}),
and one increases $\rho_-$, one will still cross the transition line at the
point given
by (\ref{cond1}). With further increase of $\rho_-$, however, the condition
(\ref{cond2})
will not be satisfied, and the system will remain in the mixed region.
Upon crossing the line  $\rho_+=1-\rho_-$, the coexistence of symmetric {it\
low-density}
regions with asymmetric ones changes into a coexistence of symmetric {\it
high-density} regions
with asymmetric ones. Finally, one ends up in the conventional high-density
phase.

One should mention that, if one analyzes the situation
more closely with second-class particles, one finds that the fraction of
symmetric
and unsymmetric configurations changes as one moves through the mixed-phase
region.
This leads to the changing `cloud' mentioned above.
However, we do not discuss this in more detail here.


\setlength{\unitlength}{2.5cm}

The complete $\rho_- - \rho_+$ -phase diagram is drawn in Fig. 10 (a-c) for
three
different values of the interaction. The sequence shows in particular how the
two
new regions develop in the upper left corner as $\nu$ is increased. In the
non-interacting case, the mixed region shrinks to the upper left part of the
diagonal
and becomes the single-chain transition line. The remaining regions
are occupied by symmetric phases with $\rho_1 = \rho_2$, and the boundaries are
determined by the extremal principle for the current $j(\rho,\rho)$, given by
(\ref{j_theor}). This happens because outside the region (\ref{ddp_domain}) the
evolution of the system at the boundaries is governed by the fast processes.
For instance, if $\rho_- > 1/2$, the injection rate will be fast. This will
produce
a considerable number of adjacent pairs of particles in the bulk and
consequently
the extraction will also be due to the fast processes.
For the symmetric boundary densities
which we consider, a symmetric bulk situation is to be expected. But then each
chain behaves essentially like an independent one with current $j(\rho) =
j(\rho,\rho)$. Consequently, the problem falls into the class considered in
\cite{Popkov99}. In drawing the boundaries, the symmetry of the problem (see
the end of section II) was used. They were also checked by simulations.

\section{Conclusion}

We have studied the problem of a two-channel system, coupled to reservoirs of
prescribed densities, through which a current flows. The aim was to see, which
phases one can expect in such a system, and what the principles are
which govern the transitions between them. The example we took, was a
one-parameter
model which has a simple stationary state on a ring. We found that already
this system shows complex behaviour in certain parts of the parameter space.

The mixed-phase region is probably the simpler phenomenon. A relatively close
analogy in equlibrium statistical physics can be found in a system of two
ferromagnetic
planes which are coupled together antiferromagnetically. The coexistence
line $H = 0, T < T_c$ of the single layers in a uniform field $H$
then widens into a whole region in the $T - H$ plane, and if one creates domain
walls
by identical boundary conditions on both layers, these tend to separate in
space,
creating bubble regions of opposite magnetizations, while a non-zero field
favours regions
of equal magnetizations. Thus one finds features as in Figs. 4 and 5.

The seesaw region with its `weak' symmetry breaking is more interesting,
and we studied it in more detail. One might view the phenomenon as a kind of
phase-
separation {\it between} the channels, as opposed to the one {\it along} the
channels in the mixed-phase region.
There are also certain similarities to critical phases. On the one hand,
the probability distribution $w(\rho)$ resembles that
for the order parameter of a finite Ising system at the critical temperature
\cite{Binder}. One the other hand, its properties do not depend
(except for the left and right limits)
on the boundary values $\rho_\pm$, a feature which it shares with the critical
(maximal-current) phase found in the present model.

One should note that our model differs from the lattice-gas models studied in
\cite{Katz,Schmitt}, since the particles cannot hop between the chains. It
would
be interesting to see whether relaxing these conservation laws changes
the situation strongly. Leaving this aspect aside, however, it seems that the
results
are rather general. We did use the microscopic details for arguing in the
paper, but the
final results do not depend on them directly. It is only the flux, which
determines the phase transition lines and characterizes the new phases. Thus
one can hope that pursuing this approach would allow to formulate generic
principles which govern the phase transitions in such multi-channel systems.

\section{Acknowledgements}

We have benefited from discussions with W. Dieterich, G. Sch\"utz and in
particular
V. Rittenberg. V. P. would like to thank the Alexander von Humboldt Foundation
for
financial support.

\appendix

\section{Boundary rates}

In our model we view the reservoirs as extensions of the
system having the same properties as the ring \cite{Antal00}.
This permits a natural definition of the boundary rates.
Consider, for example, the first process in Fig.~\ref{fig_process}. On the ring, it leads to the
average
current
\be
\langle j_l \rangle =
\al \langle n_l (1-m_l) \rangle
\langle n_{l+1} (1-m_{l+1}) \rangle
\ee
between sites $l$ and $l+1$. Here the product form of the stationary state
has been used. If site $l$ belongs to the left reservoir,
one combines $\al$ with the first factor, evaluated for the prescribed
boundary densities $\rho_1 = \rho_2 = \rho_-$. This gives the rate
$
\al_- = \al \langle n (1-m) \rangle \rho_-
$. 
Analogously, if site $(l+1)$ belongs to the right reservoir, one combines
$\al$ with the second factor and defines
$
\al_+ = \al \langle (1-n)(1-m) \rangle \rho_+
$. 

In this way, one obtains boundary rates which are determined by the bulk ones,
multiplied by the weight of the boundary configuration involved in the process.
With $\alpha = \beta = \gamma = 1$, one has
\begin{eqnarray}
&\alpha_- = \rho_- - p_- ;\ \ \beta_- = \gamma_- = p_-;\ \
\epsilon_- = \epsilon (\rho_- - p_-); \\
&\alpha_+ = \gamma_+ = p_+ +1 - 2\rho_+ ;\ \ \beta_+ = \rho_+ - p_+ ;\ \
\epsilon_+ = \epsilon (\rho_+ - p_+)
\end{eqnarray}

where $p_\pm = p(\rho_\pm) = \langle n m \rangle \rho_\pm$ and $p(\rho)$ is
given by
\be
p(\rho) = \rho + \left[ \sqrt{\left( 2\rho -1 \right)^2 -4\epsilon \rho \left(
1-\rho \right)}
-1 \right]/{2 \left( 1 -\epsilon \right)}
\ee
For $\epsilon \ll 1$, configurations with two particles (holes) at one site
have a
small weight if $\rho < 1/2$ ($\rho > 1/2$). Thus, for $\rho_- < 1/2$ and
$\rho_+ > 1/2$,
the rates $\beta_-, \gamma_-, \alpha_+, \gamma_+$ are proportional to
$\epsilon$, as
are the rates $\epsilon_\pm$. In this case, only one fast boundary process
remains at
each end.

With these quantities one can then write down the total rates for the boundary
processes. For example, if the first sites of both chains are empty, the rate
with which a particle enters one chain is given by ($\alpha_- + \gamma_-$). As
a
consequence, the currents at the left and the right end of the first chain
are, respectively
\begin{eqnarray}
&j_-^1 = \left( \alpha_- + \gamma_- \right) \left( 1-n_1 \right) \left(1-m_1
\right)
 + \left( \beta_- + \eps_- \right) \left( 1-n_1 \right) m_1 ; \\
&j_+^1 = \left( \beta_+ + \gamma_+ \right) n_N m_N + \left( \alpha_+ +
\epsilon_+ \right)
 n_N \left( 1-m_N \right);
\end{eqnarray}
If $\rho_- = \rho_+$, the stationary state of the system is automatically the
same
as for the ring, i.e. the density is constant everywhere. This is the basic
motivation for the approach described here. For $\epsilon = 1$ everything
reduces
to the single-chain problem \cite{Schuetz93,Derrida9398}.

\section{Mean-field equations}
In the mean-field approximation, one neglects the correlations
between different sites. In our case,
the system on a {\it ring} has no correlations in the stationary state
$\langle n_{l} n_{k} \rangle =
\langle n_{l}  \rangle
\langle  n_{k} \rangle $. Two adjacent sites on different chains,
however, are correlated. Therefore one should take the product $p_l=n_{l}
m_{l}$
as an independent variable.
Keeping this in mind, the mean-field equations resulting from the gain and
the loss processes become (for $\al=\beta=\gamma=1$ and leaving out the
averaging brackets for simplicity)
\begin{eqnarray}
{ \partial n_k \over \partial t}& = &j_k^1 -j_{k+1}^1 \nonumber \\
& = &\left[ n_{k-1} (1-n_{k}) + (\epsilon-1)(n_{k-1}-p_{k-1})(m_{k}-p_{k})
\right] -
\left[ n_k (1-n_{k+1}) + (\epsilon-1)(n_k-p_k)(m_{k+1}-p_{k+1}) \right]
\label{n_k} \\
{ \partial p_k \over \partial t}& = &
(n_k-p_k)
(\epsilon m_{k-1}+(1-\epsilon)p_{k-1} )+
(m_k-p_k)
(\epsilon n_{k-1}+(1-\epsilon)p_{k-1} )-
p_k(2-n_{k+1}-m_{k+1}).
\label{p_k}
\end{eqnarray}
The equation for the  $m_k$ is obtained by substituting
$m \leftrightarrow n$ in (\ref{n_k}).

The homogeneous solution of these equations
$ m_k \equiv m, n_k \equiv n,  p_k \equiv p $ leads to a quadratic
equation for $p$. When substituted into the expression
for the current in (\ref{n_k}), this reproduces the value (\ref{j_theor})
because the mean-field equations are exact in this case.

For the open system, the equations (\ref{n_k},\ref{p_k}) should be supplemented
by the boundary conditions
$n_0 = m_0 = \rho_-, n_{N+1}= m_{N+1} = \rho_+ $. For $p_k$ one takes the
homogeneous solution at both ends.


\begin{references}

\bibitem{Schmitt}
For a review, see : B. Schmittmann and R. K. P. Zia, {\it Statistical
Mechanics of Driven Diffusive Systems} in :\\ {\it Phase Transitions and
Critical Phenomena}, Vol. 17, eds. C. Domb and J. Lebowitz, Academic
Press (1995).

\bibitem{Schuetz93}
G. Sch\"utz and E. Domany, J. Stat. Phys. {\bf 72}, 277 (1993).

\bibitem{Derrida9398}
B. Derrida, M. R. Evans, V. Hakim and V. Pasquier,
J. Phys. A {\bf 26}, 1493 (1993);\\
B. Derrida, J. L. Lebowitz,  Phys. Rev. Lett. {\bf 80}, 209 (1998).

\bibitem{Rezakhanlou91} F. Rezakhanlou, Comm. Math. Phys.
{\bf 140}, 417 (1991).

\bibitem{Krug91}
J. Krug, Phys. Rev. Lett. {\bf 67}, 1882 (1991).

\bibitem{Mukamel95}  M. R. Evans, D. P. Foster, C. Godr\`eche and
D. Mukamel, Phys. Rev. Lett. {\bf 74}, 208 (1995);
J. Stat. Phys. {\bf 80}, 69 (1995).

\bibitem{Popkov99}
V. Popkov and G. M. Sch\"utz, Europhys. Lett. {\bf 48}, 257 (1999).

\bibitem{Rittenberg-chain}
P. F. Arndt, T. Heinzel and V. Rittenberg, J. Stat. Phys. {\bf90}, 783, (1998).

\bibitem{Mukamel_walker}
C. Godr\`eche, J. M. Luck, M. R. Evans, D. Mukamel, S. Sandow and
E. R. Speer,\\
J. Phys. A {\bf 28}, 6039 (1995).

\bibitem{Arndt-walker}
P. F. Arndt and T. Heinzel, J. Stat. Phys. {\bf 92}, 837 (1998).

\bibitem{Singer80}
H. Singer and I. Peschel, Z. Physik B {\bf 39}, 333 (1980).

\bibitem{Antal00}
T. Antal and G. M. Sch\"utz, Phys. Rev. E {\bf 62}, 83 (2000).

\bibitem{van Kampen}
N. G. van Kampen, {\it Stochastic Processes in Physics and Chemistry},
North-Holland (1981), Chapter XI, Equ.(2.15).


\bibitem{Binder}
K. Binder, {\it Finite-Size Effects at Phase Transitions}, in {\it
Computational Methods in Field Theory}, eds. H. Gausterer and C. B. Lang,
Lecture Notes in Physics, Vol. 409, Springer (1992).

\bibitem{Katz}
S. Katz, J. Lebowitz and H. Spohn, J. Stat. Phys. {\bf 34}, 497 (1994).

\bibitem{Derrida} B.Derrida, S.A. Janowsky, J.L.Lebowitz, E.R. Speer,
{\it J.Stat.Phys.} {\bf 73}, 813 (1993).

\bibitem{local_decrease} The local decrease of the average density
at the left end of the bubble is an artefact of the chosen rates.
Particle $A$ tends to have empty sites in both chains immediately
to its left.

\bibitem{Rittenberg_thanks} We are indebted to V. Rittenberg for pointing this out.


\end{references}
\end{document}